\documentclass[runningheads]{llncs}
\usepackage[T1]{fontenc}
\usepackage{graphicx}
\usepackage[bookmarks,colorlinks=true,allcolors=blue,pdfhighlight=/O]{hyperref}
\usepackage{color}

\usepackage[nameinlink]{cleveref}
\usepackage[dvipsnames]{xcolor}

\begin{document}
\title{Digital Twins for Small Towns and Rural Regions: Data Integration, Simulation and Visualisation Across Three Use Cases in Lower Austria}
\titlerunning{Digital Twins for Small Towns and Rural Regions}
%
\author{Thomas Delissen\inst{1}\orcidID{0009-0005-1769-9449} \and
Lukas Daniel Klausner\inst{1}\orcidID{0000-0003-3650-9733} \and
Maximilian Lackner\inst{1}\orcidID{0009-0004-9040-2982} \and
Torsten Priebe\inst{1}\orcidID{0000-0001-9282-2535} \and
L\H{o}rinc Thurnay\inst{2}\orcidID{0000-0003-2374-921X}}
\authorrunning{T. Delissen et al.}
%
\institute{University of Applied Sciences St. Pölten, St. Pölten, Austria
\and
University for Continuing Education Krems, Krems an der Donau, Austria}
\maketitle              
\begin{abstract}
    Smart city and digital twin concepts hold considerable potential for improving local governance and planning, yet practical implementations have almost exclusively focused on large metropolitan areas. Small towns and rural regions face a distinct set of challenges – limited data availability, constrained budgets and lower digital capacity – that make a direct transfer of urban approaches infeasible. This article presents results from the ongoing project ``Smart Cities and Digital Twins in Lower Austria'', which adapts digital twin concepts for small-town and rural contexts following a design science research approach. Based on stakeholder workshops with municipalities in Lower Austria, three use cases were identified and implemented as prototype digital twins: traffic flow planning (emphasising data integration), population growth (emphasising simulation) and shared spaces (emphasising visualisation). We describe the design and implementation of each prototype and report on a technical evaluation of their feasibility. Our findings highlight both the potential and the specific limitations of digital twins in resource-constrained settings and offer practical guidance for municipalities, policymakers and researchers working in non-urban contexts.
    
    \keywords{digital twin \and small cities and towns \and smart city \and smart region \and simulation}
\end{abstract}

\setcounter{footnote}{0}

\section{Introduction}
A smart city leverages new technologies, such as information and communication technologies, to manage complex societal challenges and enhance residents' quality of life across a variety of dimensions, including economic improvement, social stability and environmental sustainability~\cite{VialePereiraTemple2024}. As the digital transformation of urban environments accelerates, digital twins have emerged as powerful tools to support policymakers in simulating scenarios and developing smart, sustainable solutions~\cite{PereiraKlausner2023}. However, these concepts have thus far been predominantly implemented in large metropolitan areas, such as Barcelona or Milan~\cite{GascoTrivellato2016}. Transferring these solutions to small towns and rural regions is not a straightforward process -- these non-urban areas face distinct challenges, including constrained financial resources, limited data availability and lower digital capacity, which collectively hinder the adoption of complex digital twin technologies~\cite{SevcikChaloupkova2023}. Furthermore, a one-size-fits-all approach is insufficient, as the societal and demographic realities of rural areas often differ starkly from those of large cities.

The study presented in this article is part of the project ``Smart Cities and Digital Twins in Lower Austria'', which seeks to bridge this gap by adapting smart city and digital twin approaches to smaller-scale urban and non-urban contexts. It focuses on using digital twins to improve the understanding of the interaction between smart sustainable solutions and city inhabitants, thus supporting evidence-based policymaking and sustainable local governance. It is conducted in the state of Lower Austria, a region characterized by a lack of major urban centers (its capital and largest town has approximately 60,000 inhabitants), but a strong political commitment to the digitalization of the public sector and the promotion of regional smart initiatives~\cite{noe_digireport_2023}.

Designing digital twins for small towns and rural regions requires simple, sustainable and reusable solutions that remain feasible under limited data, budgets and technical expertise. To ensure local relevance, such systems should be co-developed with stakeholders and practitioners who understand their communities’ specific needs. In our project, we therefore adopted a transdisciplinary approach and conducted workshops with representatives from municipalities in Lower Austria to identify pressing local challenges and assess the feasibility and usefulness of potential digital twin applications~\cite{TempleKaltenbrunner2025}. These collaborative sessions led to the selection of three use cases for further development: traffic flow planning, population growth and shared spaces. Because abstract discussion of such use cases often makes it difficult for non-technical stakeholders to grasp the functionality and practical benefits of digital twins, we implemented prototype digital twins for each scenario. This article describes how these prototypes were co-designed, how they work and how their technical feasibility and usefulness were evaluated under the specific constraints of non-urban environments.

Our contributions are as follows:
\begin{itemize}
    \item We demonstrate that digital twin prototypes for small towns are technically feasible when scoped as minimum viable products around data integration, simulation and visualisation.
    \item We present three concrete prototype implementations (traffic flow planning, population growth and shared spaces) developed through stakeholder co-design in Lower Austria, with full technical detail.
    \item We characterise the key barriers to scaling these prototypes into production-ready systems, providing actionable guidance for municipalities, policymakers and researchers in resource-constrained settings.
\end{itemize}

\section{Background and Methodology}\label{sec:background}
This work follows a design science research (DSR) paradigm in information systems (IS). DSR focuses on the purposeful design and evaluation of artefacts that address relevant real-world problems. In IS research, such artefacts may take the form of constructs, models, methods or instantiations~\cite{hevner2004design}. In this study, the artefacts are three prototype digital twins developed as minimum viable products (MVPs) for selected smart city use cases in small towns and rural regions.

To structure the research process, we draw on the design science research methodology (DSRM) by Peffers et al.~\cite{peffers2007design}. This process model fits the present study well: Relevant challenges and use cases were identified with stakeholders; prototype artefacts were developed for the selected scenarios; and these artefacts were then evaluated with regard to their feasibility and usefulness.

The process is iterative rather than linear, as insights from development and evaluation may lead to revised solution ideas or even to reframing the problem. This iterative character was particularly important in the present work, where the design of digital twin prototypes required repeated adaptation to data constraints, simulation requirements and visualisation needs. We present our methodology by aligning the steps of our research with the five phases of the design science research cycle following Vaishnavi and Kuechler, Jr.~\cite{vaishnavi2015design}:

\begin{itemize}
    \item In the \emph{Awareness of Problem} phase, the present research builds on prior transdisciplinary work with municipalities and public administration stakeholders in Lower Austria~\cite{TempleKaltenbrunner2025}. The problem addressed in this work is therefore how digital twin artefacts can be designed for small-town and rural contexts in a way that remains feasible under data and resource constraints while still supporting decision-making, simulation and communication. (For a brief discussion of these constraints and other related work, see \Cref{sec:relwork}.)
    \item In the \emph{Suggestion} phase, these use cases were translated into a tentative design. Rather than pursuing one comprehensive digital twin, we proposed three focused prototypes, each emphasising one key challenge dimension for small towns and rural regions: data integration, simulation and visualisation. This made the selected use cases operationalisable in a feasible yet still informative manner.
    \item In the \emph{Development} phase, the tentative design was implemented as three digital twin prototypes (MVPs). The emphasis was not on technical novelty in the implementation itself, but on constructing artefacts that instantiate the proposed design in a feasible and usable form. Accordingly, the use cases were realised with state-of-practice technologies as presented in \Cref{sec:implementation}.
    \item In the \emph{Evaluation} phase, the artefacts are assessed against the expectations formulated in the problem awareness phase, namely whether digital twin prototypes for the selected use cases can be implemented in a feasible and useful way. \Cref{sec:evaluation} presents a preliminary technical evaluation along the dimensions of data integration, simulation and visualisation. The detailed stakeholder evaluation is beyond the scope of this article.
    \item In the \emph{Conclusion} phase, the results of the design research effort are consolidated and the knowledge gained from the construction and evaluation of the artefacts is reflected upon. In the present work, this means summarising what can be learned from the prototypical implementation of digital twins for small towns and rural regions. \Cref{sec:conclusion} concludes the article and discusses future work; in particular, a full conclusion of the DSR cycle will be provided based on the ongoing stakeholder evaluation in a follow-up publication.
\end{itemize}

\section{Related Work}\label{sec:relwork}
The conceptualisation of smart cities extends beyond the mere deployment of new technologies, relying heavily on governance frameworks that foster collaboration among local authorities, the private sector and citizens~\cite{MisuracaLipparini2021}. While early smart city strategies primarily emphasised urban networks, recent literature highlights the necessity of an IS perspective to fully grasp how information and communication technologies (ICT) can improve residents' quality of life and sustainable resource management~\cite{IsmagilovaHughes2019,VialePereiraParycek2018}. This IS-centric approach underscores the importance of transparent, collaborative decision-making~\cite{VialePereiraEibl2018,VialePereiraMacadar2017}. Furthermore, the digital transformation of public administration requires identifying localised needs rather than relying on generalised, one-size-fits-all solutions~\cite{CaragliuDelBo2019,MergelEdelmann2019}.

Complementing the broader smart city paradigm, digital twins have emerged as pivotal tools for urban management. Defined fundamentally as virtual representations of physical systems that continuously exchange information with their real-world counterparts~\cite{VanDerHornMahadevan2021}, digital twins enable policymakers to simulate complex behaviours and predict the outcomes of policy decisions~\cite{AlamElSaddik2017,ShiPan2023}. Traditionally, the deployment of such technologies has been concentrated in large, densely populated metropolitan areas where extensive data collection is already established~\cite{WangChen2023}. In these environments, digital twins offer significant benefits, ranging from cost reductions to optimised asset lifecycle management~\cite{HuNguyen2018,MacchiRoda2018}.

However, adapting digital twin concepts for small towns and rural regions introduces both unique challenges and opportunities. Unlike major urban centers, non-urban usage contexts often face a scarcity of real-time data, complex environmental interdependencies and a need for highly context-specific modelling~\cite{NooriHoppe2023,TrantasPlug2023}. Recent research has begun to address these specific gaps by implementing proof-of-concept prototypes that strongly resonate with the core challenges addressed in our own work: data integration, simulation and visualisation. Addressing the data integration hurdle, Zhang and Zhao~\cite{ZhangZhao2026} proposed collaborative frameworks designed to overcome information silos by streamlining the collection and processing of heterogeneous data sources tailored specifically for rural environments. The application of such integrated data often necessitates focusing on distinct, high-impact use cases, particularly concerning mobility and demographic shifts. In the domain of transport infrastructure, dedicated simulation tools have been successfully employed to model traffic dynamics and improve road networks in small towns~\cite{NikolaevaSakhapov2020}. This directly aligns with our traffic flow planning prototype in \Cref{subsec:trafficflowuc}, which similarly simulates the effects of new road infrastructure, such as a bypass, on local traffic patterns to aid municipal decision-making. Beyond physical infrastructure, modelling human factors and fluctuating demand through advanced simulation is equally critical. Fendrich et al.~\cite{FendrichJanzso2024} proposed multi-layered, agent-based digital twin frameworks to capture individual decision-making and demographic dynamics, aiming to support sustainable rural development. Emphasising the need for cost-effective mobility and resource solutions in non-urban settings, Bae et al.~\cite{BaeBerndt2024} further explored real-time digital prototypes to optimise cooperative transport. These efforts highlight the importance of simulating capacity, demand, and human behaviour a challenge we similarly address in our population growth prototype in \Cref{subsec:populationgrowthuc}, which models the allocation of municipal services in response to demographic changes.

Finally, to make these simulations and urban redesigns tangible for local stakeholders, the extraction and synthesis of spatial data for accurate visualisation have proven essential. Blagoni'{c} et al.~\cite{BlagonicMarkovinovic2025} demonstrated how the integration of diverse spatial datasets and detailed 3D modelling can effectively support urban regeneration in small historic towns. Their approach of combining existing spatial records with 3D data collection techniques closely parallels the methodology behind our prototype for visualising shared spaces in \Cref{subsec:sharedspacesuc}, which relies on comparable 3D modelling strategies to make urban transformations comprehensible. Crucially, across all these applications, digital twins must serve as accessible communication tools. By establishing system boundaries that reflect local demographics~\cite{AppioLima2019} and designing interactive, user-friendly interfaces, digital twins can foster transparent public engagement and ensure that simulated policies align with actual resident needs from the very beginning of the design process~\cite{AbdeenShirowzhan2023,GengDu2021,RamuBoopalan2022}.

\section{Implementation of the Digital Twin Prototypes}\label{sec:implementation}
During the analysis of the first, broader array of fifteen use cases that were derived during the project's requirements phase~(for details, see~\cite[Table~II]{TempleKaltenbrunner2025}), it became clear that most use cases shared some key commonalities. In particular, we identified three main aspects or characteristics of a digital twin that appeared to be crucial for most use cases:
\begin{itemize}
    \item data integration, to ensure the digital twin adequately represented reality;
    \item simulation of scenarios, to answer “what-if” scenarios and provide practically useful output; and
    \item visualisation and interaction, to allow users, especially laypeople, to engage with the digital twin.
\end{itemize}

\noindent For each of the use cases, we assessed the feasibility of data integration, simulation and visualisation, and found that feasibility varied considerably depending on the case. While some scenarios proved impractical because of missing data or the difficulty of modelling complex natural or human behaviour, visualisation was generally the most feasible aspect thanks to the availability of mature open-source tools for both 2D and 3D applications. As explained in detail in \cite[Sections~II and~III]{TempleKaltenbrunner2025}, we narrowed the fifteen use cases down to three for which we created digital twin prototypes as minimum viable products (MVPs for short) -- the ``artefacts'' in our design science research (see \Cref{sec:background}). 

To further evaluate the impact and complexity of the three aspects (data integration, simulation and visualisation), the three use cases each corresponded with one of the three aspects with higher relevance for that specific use case: data integration was a particular challenge for the traffic flow planning MVP; the population growth MVP required a focus on simulation; and, quite naturally, the shared spaces MVP was most strongly connected to visualisation. We now describe each of the three MVPs we developed in detail before describing the process and results of our first, technical evaluation phase.

\subsection{Traffic Flow Planning Use Case}\label{subsec:trafficflowuc}
Small towns and rural municipalities are dynamic environments where demographic shifts, newly developed residential areas, and evolving local businesses continuously alter movement and traffic patterns. Understanding these dynamics is crucial for urban planners and policy-makers to optimise traffic flows, alleviate congestion, and minimise environmental impacts. For instance, while a strategically placed bypass can significantly reduce bottlenecks in a town centre and improve residents' quality of life, poor planning may inadvertently lead to induced demand or fail to reduce existing congestion.

The traffic flow planning MVP addresses this challenge by providing a simulation that models the effects of a new bypass in a small Austrian municipality. This prototype offers an interactive and accessible illustration of how infrastructure changes impact existing traffic flows. Beyond supporting internal municipal decision-making, such simulations serve as vital communication tools to engage stakeholders and the broader public in the planning process. Crucially, because this specific use case was selected to foreground the challenge of \emph{data integration}, the MVP leverages historical traffic data from a recently constructed bypass to demonstrate how real-world data can be effectively integrated into a digital twin environment. 

\begin{figure*}[t!]
    \centering
    \begin{minipage}{0.48\textwidth}
        \centering
        \includegraphics[width=\linewidth]{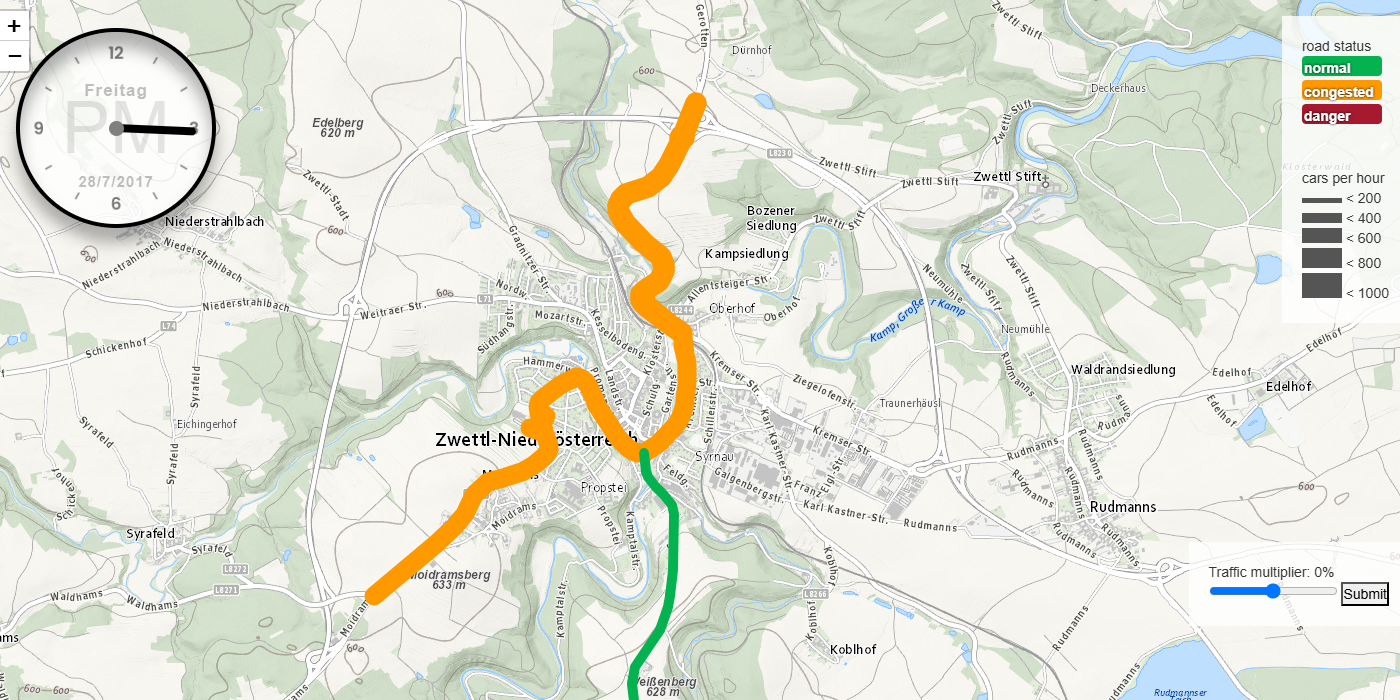}
    \end{minipage}
    \hfill
    \begin{minipage}{0.48\textwidth}
        \centering
        \includegraphics[width=\linewidth]{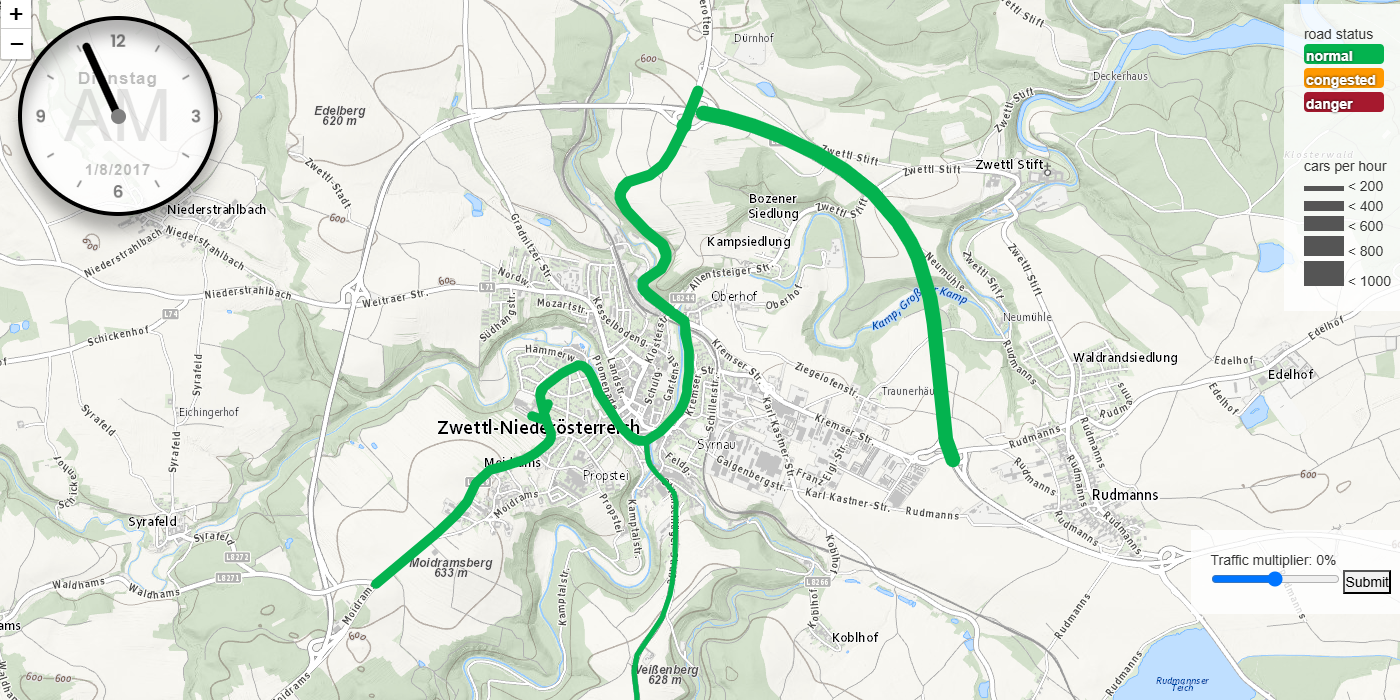}
    \end{minipage}
    \caption{User interface of the traffic flow planning MVP, showing congestion patterns before (left) and after (right) the bypass was opened. The simulation demonstrates that the new bypass (green line, right) reduces the absolute hourly traffic volume on the road through the city, shown by the thinner line. Consequently, the road's status improves from ``congested'' to ``normal'', as indicated by the colour change.}
    \label{fig:bypass}
\end{figure*}

For this MVP, data from the traffic department of the Lower Austrian state government were used that showed hourly traffic counts on streets in and around the city of Zwettl (population slightly above 10,000 people), from before and after a bypass was built. The data were provided in tabular form, describing traffic counts from multiple locations in and around Zwettl. Based on these data, the MVP could show how the traffic load changed after the opening of the B38 bypass. In the simulation, the traffic is visualised for the period between 2017-07-28 and 2017-08-02. These dates were chosen because the B38 bypass was opened for public access on 2017-08-01 and thus the actual impact on traffic flows caused by the new bypass can be shown. The data clearly show a significant reduction of traffic through Zwettl from the hour that the bypass opened.

The simulation highlights the roads for which measurement data were available on a two-dimensional base map. The roads' width represents the aggregate amount of traffic for a specific hour and a colour gradient indicates whether that traffic load is ``normal'', ``congested'' or ``danger[ous]'' (see \Cref{fig:bypass}). The category thresholds were chosen for illustrative purposes; more domain expertise would be required to calibrate this part of the MVP. Some parts of the traffic flows are adjustable by the user (e.\,g.\ by applying a traffic multiplier to simulate busier or quieter days).

\subsection{Population Growth Use Case}\label{subsec:populationgrowthuc}
The demographic landscape of small towns is subject to continuous change. Municipalities must proactively adapt, as population growth directly increases demand for essential civic services such as housing, childcare, geriatric care and utility infrastructure.\footnote{Conversely, population decline requires the careful reallocation of shrinking resources. For the sake of clarity, we will henceforth refer only to population \emph{growth}.} The objective of this use case is to design a digital twin that simulates demographic scenarios to help local governments anticipate and manage these shifts.

\begin{figure*}[t!]
    \centering
    \begin{minipage}{0.48\textwidth}
        \centering
        \includegraphics[width=\linewidth]{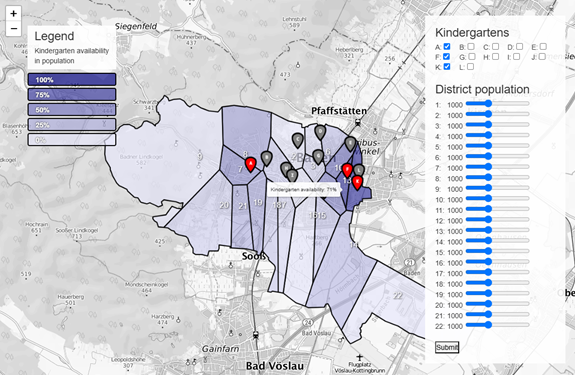}
    \end{minipage}
    \hfill
    \begin{minipage}{0.48\textwidth}
        \centering
        \includegraphics[width=\linewidth]{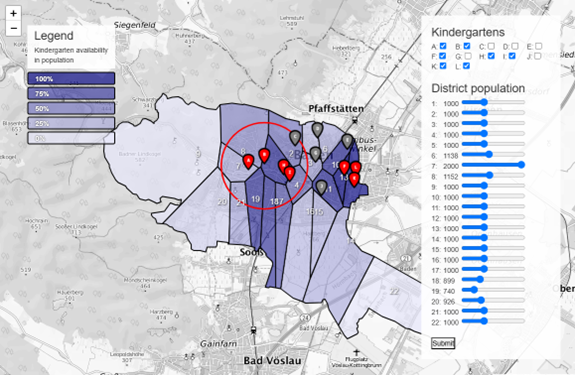}
    \end{minipage}
    \caption{User interface of population growth MVP. Red dots mark kindergartens that are currently open; the colour gradient for the areas shows the service coverage. In the left image, only three locations are open; the right image shows improvements in coverage through the darker gradient for most areas. The right image shows two more features of the MVP: Populations can be adjusted individually for each area using the sliders and the range of the currently selected location B is visualised with a red circle.}
    \label{fig:population}
\end{figure*}

The resulting MVP operates as a supply-and-demand system: As the simulated population grows, demand for municipal services (such as kindergartens) rises accordingly. By modelling these dynamics, the digital twin supports evidence-based decisions on long-term infrastructure investment and service allocation. Because the core challenge of this use case is \emph{simulation}, the implementation prioritises the underlying computational models over complex visual representation. A simple interactive two-dimensional map was therefore considered sufficient to communicate the results effectively.

For this MVP, the town of Baden bei Wien (with a population of about 25,000 people) was selected. While the MVP is limited to a single service (kindergartens), the concept can readily be extended to others. On a simple map of Baden bei Wien, the location of each kindergarten is shown; the municipality is divided into a limited set of areas to provide an intermediate degree of pooling and grouping of population data. Each area is colour-coded along a gradient to indicate how well it is served by the existing kindergartens (see \Cref{fig:population}). Users can enable or disable kindergartens to simulate the effects of specific facilities’ operational status on service coverage. They can also adjust the population of each area, simulating future growth or decline. These simple controls allow a range of scenarios to be explored; for example, adding a new kindergarten at different locations can be compared when looking for the best option, and demographic scenarios can be tested to identify future coverage challenges as populations shift across areas.

The areas are based on the polling areas (``Wahlsprengel'') for the 2024 European Parliament elections. Although this may seem somewhat unorthodox, there are no official permanent subdivisions of suitable granularity; polling areas also have the advantage of being roughly uniform in population. We applied some simplifications to the data, as the focus of this use case was simulation rather than data integration: Polling areas are defined by address ranges, and coordinates for the corresponding street names were obtained from OpenStreetMap. Some manual data cleaning was required to resolve discrepancies in street name spellings and issues with overlapping areas.

For the base scenario, the population of each of the 22 areas was set to 1,000, resulting in a total population of 22,000 and thus slightly undercounting the actual population at the time of our study. These numbers can be adjusted by the user, allowing domain experts to enter more accurate figures instead. To keep the MVP simple, the capacity of each kindergarten was fixed at 1,500; this represents the capacity to serve the entire population of an area, not only children (that is, we assume one kindergarten can cover the childcare needs of 1,500 residents with children aged six years or younger).

The simulation distributes kindergarten capacity across areas based on geographic proximity using a Monte Carlo method. The closer an area's geographic centre is to a kindergarten, the more likely it is that a capacity unit is assigned to that area. Areas cannot exceed 100\% capacity; if an area is fully covered but contains kindergartens with unused capacity, the surplus is distributed to other areas within reach. The range of each kindergarten is defined as approximately 1.15 kilometres; again, this figure was chosen only to demonstrate the MVP's capabilities, and for actual applications it would likewise be configurable. This MVP is built on the same software stack as the traffic flow planning use case, allowing the reuse and combination of both implementations.

\subsection{Shared Spaces Use Case}\label{subsec:sharedspacesuc}
Main streets in Austrian municipalities have traditionally been designed with a strict segregation of traffic modes, predominantly favouring motorised vehicles while relegating pedestrians and cyclists to designated pavements and lanes. Recently, many local governments have considered redesigning central streets into ``shared spaces'', lifting these strict demarcations so that cars, cyclists and pedestrians coexist in a single, unified area. This approach typically enforces slower driving speeds and heightened driver awareness, aiming to improve overall safety, enhance the pedestrian experience and revitalise local communities.

\begin{figure*}[t!]
    \centering
    \begin{minipage}{0.32\textwidth}
        \centering
        \includegraphics[width=\linewidth]{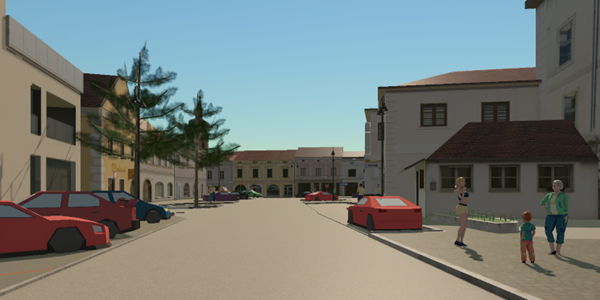}
    \end{minipage}
    \hfill
    \begin{minipage}{0.32\textwidth}
        \centering
        \includegraphics[width=\linewidth]{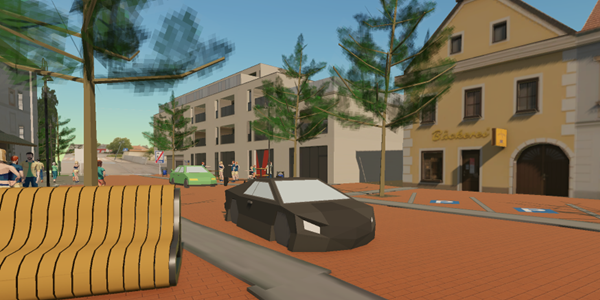}
    \end{minipage}
    \hfill
    \begin{minipage}{0.32\textwidth}
        \centering
        \includegraphics[width=\linewidth]{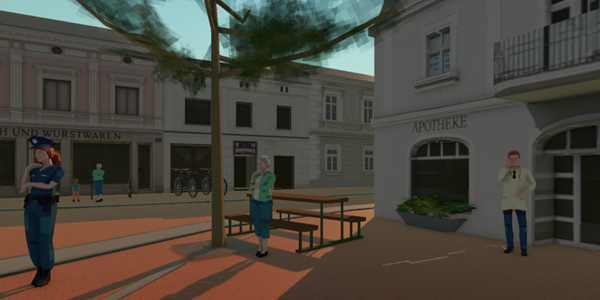}
    \end{minipage}
    \caption{Views from the shared space MVP. The image at the left shows the street in its current state; the centre image shows a possible redesign of the street layout as a shared space. The image at the right shows that the MVP already features some customisation, such as changing the time of day.}
    \label{fig:sharedspace}
\end{figure*}

However, urban transformation projects often encounter significant public resistance. Stakeholders frequently voice concerns that shared spaces might impede traffic flow, negatively affect local retail businesses or inadvertently compromise the safety of vulnerable populations. Because it is inherently difficult for the public to envision such structural changes, the shared spaces MVP was designed specifically to tackle the challenge of \emph{visualisation}. By providing a realistic, interactive 3D digital twin of a proposed street redesign, the MVP offers a tangible basis for informed dialogue among stakeholders. Visualising the future state in this manner helps to bridge the imagination gap, fostering constructive public engagement and mitigating resistance to change. 

For our MVP, a specific street in a municipality our stakeholders were familiar with was chosen: the main square in Traismauer (with a population of about 6,000 people). The current situation of the street was modelled as well as what the street could look like if it were a shared space. Using OpenStreetMap, the street's geographical information was extracted, including the exact shapes of the buildings on that street. However, it was not possible to find detailed information on the pavement layout, building textures or the street's surface. We hence went out into the street and took pictures of all the buildings as well as of specific design elements, such as the tiling on the pavement, flowerbeds and other details. These data were combined with screenshots from Google Street View\footnote{\url{https://www.google.com/streetview/}} to provide information for the MVP's baseline scenario.

Based on the photo data, we extracted the relevant image regions and used Adobe Photoshop's inpainting feature to generate usable textures for the buildings along the street as well as for the pavement and street. We were able to create textures that were quite close to the real-world ones, but slightly normalised and simplified; for example, shadows and dirt visible in the original photos were removed so the textures could be used in 3D modelling tools. Textures from the buildings together with the shapefiles from OpenStreetMap were then imported into the 3D modelling tool Blender.\footnote{\url{https://www.blender.org/}} We then used Blender to create 3D models of all the buildings, streets and pavements, with realistic textures and dimensions matching the real data from OpenStreetMap. The output was a static 3D model of the street and its buildings. 

In the next step, the models were loaded into Unity,\footnote{\url{https://unity.com/}} a game engine we used to create the interactive experience required for the MVP. To save time, benches, trees, bushes and street signs were not manually modelled. We instead used so-called ``prefabs'' for these, i.\,e.\ 3D assets provided under free or lenient licences. These prefabs were not exact matches for the physical objects in Traismauer, but were deemed close enough for our purposes. We included two types of dynamic elements, cars and pedestrians. For these, animated prefabs were used, and we wrote manual scripts to move these objects through the street in simple movement cycles.

After recreating the current street layout, we created a complete redesign of the streets and pavements to demonstrate what a shared space at this location might look like (see \Cref{fig:sharedspace}). We based the shared space layout and design decisions on recent shared space projects from Vienna and a variety of Lower Austrian cities; again, the focus for this MVP lay on visualisation, not on data integration or a high degree of realism. We also implemented a variety of interactive features, allowing users to freely walk through the street and look around; switch between the current street design and a shared space redesign; toggle between different numbers of cars in the parking spaces; and select different times of day (with different lighting conditions).

\section{Technical Evaluation}\label{sec:evaluation}
The comprehensive evaluation of the MVPs with stakeholders from small towns and rural areas in Lower Austria is currently ongoing; for this article, we will present preliminary results with a focus on the technical aspects of the prototype implementations. To evaluate the feasibility of each use case, we evaluated the data integration, simulation and visualisation needs that were required to create the MVPs and extrapolated this to the efforts that would be required to create a fully functional digital twin for each use case. We also identified several technical challenges that would complicate the implementation of such digital twins. 

\textbf{Data integration.} For the traffic flow planning MVP, existing data from Lower Austria were used, which means we visualised the actual traffic changes that occurred when the bypass was implemented. This is different from how a fully functional digital twin for this use case would work, because the digital twin is needed \emph{before} the bypass is planned and built. This means that the traffic effects from the bypass need to be estimated, modelled or simulated. Also, traffic data are not available for all streets; this means that even though it is possible to show the changes of traffic on the main streets, for smaller side streets, estimates need to be made.

For the population growth MVP, the data for the current state of a municipality at large are available in the form of public open data. In Europe, most countries provide detailed information about each municipality online which can be used to model the current situation. However, breaking down this information onto smaller areas can be hard as this information often is not publicly available. Moreover, the capacity of different services (such as schools, hospitals or police departments) is usually not easy to estimate. Therefore, such a digital twin cannot be created fully based on open data. Assumptions and approximations need to be made, which negatively affect the usefulness of the digital twin. 

For the shared spaces MVP, open data about street layouts are available and can be used to create a first 3D representation of the street. However, detailed information (such as building or street and pavement textures) is harder to extract; for that, manual data collection is necessary. How many cars, bicycles and pedestrians are present in the current street situation can, in theory, be estimated based on e.\,g.\ cell phone data. However, these data are not open and will usually be too expensive for small municipalities to use as a data source.

\textbf{Simulation.} As already mentioned, for the traffic flow planning MVP, the impact of creating a bypass on traffic patterns needs to be simulated, which is a difficult and context-dependent task that is different for each town or region and can therefore not be simulated universally. This means that creating a digital twin for this use case would require municipalities or state governments to estimate the impact of a bypass with current means; a digital twin would mainly function as a visualisation tool in this case.

For the population growth MVP, simulation is the most important aspect. The main simulation engine implemented in the MVP is mainly calculating differences between supply against demand for certain services, based on population size. The challenge with this is not so much the simulation itself but getting the numbers right: How many people can a kindergarten support and how does this capacity change over time? The stopgap that was used in the MVP to overcome this challenge was to make all relevant numbers configurable. A domain expert could hence enter the numbers for both capacity and demand in the tool to simulate and visualise the difference in line with actual expert knowledge.

In the shared spaces MVP, ideally one would like to simulate how the shared space redesign would affect the traffic and movement patterns in the street. However, data about current traffic patterns are usually not available as open data and it is also difficult to predict what the impact of a specific shared space concept is. Therefore, simulating these parts of the scenario requires conducting specific context-dependent research for the street under consideration, which is likely to be prohibitively expensive for most small towns. 

\textbf{Visualisation.} For both the traffic flow planning MVP as well as the population growth MVP, an interactive web-based map was used as the foundation for a digital twin. Such a visualisation is relatively easy to create with modern web technology, so from this perspective, the visualisation aspect is actually the easiest challenge to solve. Also, there is potential for reuse: The visualisation engine created for these MVPs is not limited to a specific city; it can work for any region in Austria, as long as the data are available for that region.

In contrast, this is not the case for the 3D visualisation that was used for the shared spaces MVP. Creating this MVP involved a substantial amount of manual labour; while the building shapes can be automatically generated based on OpenStreetMap data, creating realistic textures, modelling the actual street and pavement layouts etc.\ cannot easily be automated or transferred between different locations. Further research is needed to explore how to make this kind of digital twin scalable and reusable.

\section{Conclusion and Future Work}\label{sec:conclusion}
This article presented three prototype digital twin implementations – covering traffic flow planning, population growth and shared spaces. Following a design science research methodology, the prototypes were scoped as minimum viable products, each foregrounding one of three core implementation challenges: data integration, simulation and visualisation. A technical evaluation of the prototypes showed that web-based, interactive visualisation is the most accessible of the three dimensions, with reusable components generalisable across municipalities. Simulation of supply-and-demand scenarios, as demonstrated in the population growth prototype, is feasible but depends critically on expert-calibrated input parameters that are rarely available as open data. Data integration for traffic modelling proved the most demanding challenge, as high-quality, localised traffic data are largely unavailable to small municipalities without dedicated infrastructure or state-level support.

The stakeholder evaluation of the prototypes with representatives from municipalities and the state government in Lower Austria is currently ongoing and will be reported in a follow-up publication, together with a fuller discussion of how the prototypes were received, what modifications stakeholders requested and what lessons can be drawn for co-designing digital twins with non-technical users in non-urban settings. Alongside this evaluation, we are drafting a general architectural design for a digital twin platform tailored to the constraints of small towns and rural regions, grounded in principles derived from both our technical experience with the prototypes and from stakeholder input – briefly illustrated here – which will be elaborated in detail in that same future work:
\begin{itemize}
    \item Use open data, to avoid procurement costs and ensure universal access.
    \item Use open-source technology, to keep operational costs low.
    \item Be as simple as possible to implement, to minimise implementation costs.
    \item Support web-based interaction, to provide broad and barrier-free access.
\end{itemize}

\noindent Beyond the platform architecture, we are also developing a practical roadmap for digital twin adoption in small towns and rural regions, drawing directly on the lessons learned from the three MVP implementations described in this article.

Taken together, we hope this work contributes to making digital twin technologies genuinely accessible to the many municipalities that lack the resources of large cities, yet stand to benefit considerably from better data-driven tools for planning and governance.

\begin{credits}
\subsubsection{\ackname} This research was funded by the Gesellschaft für Forschungsförderung Niederösterreich (GFF NÖ) project GLF21-2-010 ``Smart Cities and Digital Twins in Lower Austria''. The financial support by the Gesellschaft für Forschungsförderung Niederösterreich is gratefully acknowledged.
We are also grateful to all interview and workshop participants for sharing their knowledge, experiences and insights.
\vspace*{-10pt}
\subsubsection{\discintname}
The authors have no competing interests to declare that are relevant to the content of this article.
\end{credits}
\vspace*{-5pt}
%
%
%
\bibliographystyle{splncs04}
\bibliography{Bibliography}

\end{document}